\documentclass[10pt]{iopart}
\usepackage[]{graphicx,amssymb}
\begin{document}



\title{Electronic structure of amorphous germanium disulfide via density functional molecular dynamics simulations}

\author{S\'ebastien Blaineau and Philippe Jund$^\star$}
\address{
Laboratoire de Physicochimie de la Mati\`ere Condens\'ee , 
Universit\'e Montpellier 2, \\Place E. Bataillon, Case 03, 
34095 Montpellier, France
}

\begin{abstract}
Using density functional molecular dynamics simulations we study the electronic properties of glassy g-GeS$_2$. We compute the electronic density of states, which compares very well with XPS measurements, as well as the partial EDOS and the inverse participation ratio. We show the electronic contour plots corresponding to different structural environments, in order to determine the nature of the covalent bonds between the atoms. We finally study the local atomic charges, and analyze the impact of the local environment on the charge transfers between the atoms. The broken chemical order inherent to amorphous systems leads to locally charged zones when integrating the atomic charges 
up to nearest-neighbor distances.
\end{abstract}
\pacs{PACS numbers: 61.43.Bn,61.43.Fs,71.15.Pd,71.23.-k}
 
\section{INTRODUCTION}

Germanium binary chalcogenide glasses are of great interest 
because of their specific physical properties. For example, amorphous 
GeS$_2$ systems present promising applications as high efficiency 
optical amplifiers in optical communications networks, memory or switching 
devices \cite{kolomiets,ovshinsky}, high-resolution inorganic photoresistors, 
or anti-reflection coatings \cite{malek}. Moreover, thin films doped with 
silver are promising materials for submicrometer lithography\cite{hugget,leung}, and bulk glasses with Ag$^+$ cations are good solid electrolytes with a 
high ionic conductivity at room temperature \cite{robinel,tranchant}.
 Therefore, the structural and electronic properties of g-GeS$_2$ have been 
extensively studied for many years through experimental measurements 
\cite{exp1,exp2}, and recently {\em ab initio} cluster calculations of 
crystalline GeS$_2$ have been performed on this topic \cite{foix}. However, 
in order to analyze rigorously the electronic structure of such amorphous 
materials (taking into account all the structural defects) Molecular Dynamics (MD) simulations on extended systems 
may be a particularly interesting tool since they give access to an atomic
description of the electronic properties. The structural disorder of these 
glasses has been studied in detail in a previous work \cite{blaineau2}, 
showing the evidence of bond defects and homopolar bonds (Ge-Ge and S-S) 
in glassy GeS$_2$.\\
Here we propose a theoretical study of the corresponding electronic structure 
of amorphous GeS$_2$ via a Density Functional Theory (DFT) based MD 
program called {\sc{fireball}} \cite{fireball}, which has been used 
previously for the study of analogous chalcogenide glasses and which has 
given excellent results \cite{drabold,junli}.\\
The paper is organized as follows: In Sec. II we briefly present the 
theoretical foundations of the model. In Sec. III the results are
presented and analyzed in detail in two subsections: first the 
standard structural configurations are considered, second the defects in our glassy sample are examined. A third subsection deals with the study of the local atomic charges depending on the structural environment of the atoms. Finally, 
in Sec. IV we summarize the results and give the major conclusions.

\section{THEORETICAL FOUNDATIONS}
The model we have used was developed by Sankey and Niklewski \cite{sankey}. It is included in a first-principles type molecular dynamics code, called {\sc{fireball96}}, that is based on the DFT \cite{hohenberg-kohn}, within the Local Density Approximation (LDA) \cite{ceperley,perdew}. 
A tight-binding-like linear combination of pseudoatomic orbitals, satisfying the atomic self-consistent Hohenberg-Kohn-Sham equations \cite{kohn-sham}, is used to determine the electronic eigenstates of the system. A basis set of one $s$ and three $p$ pseudo-orbitals, slightly excited to vanish outside a cut-off radius of $5a_o$ (2.645 \AA), is required.\\
The pseudopotential approximation is used to replace the core electrons by an effective potential which acts on the valence electrons, and Hamman-Schluter-Chiang pseudopotentials are used \cite{schluter}. The Hamiltonian is calculated through the Harris functional \cite{harris}, which avoids long self-consistent calculations by a zero$^{th}$ order approximation of the self-consistent density. More details on the method can be found in the original paper \cite{fireball}. \\
The pseudo-wave function $\Psi$ of the system is given by the following equation:
\begin{eqnarray}
\Psi_j(\vec{k},\vec{r})=\Sigma_{\mu}C_{\mu}^j(\vec{k})\Phi_{fireball}^{\mu}(\vec{r})
\end{eqnarray}
where j is the band index, $\Phi_{fireball}^{\mu}$ is the fireball basis function for orbital $\mu$, and C$_{\mu}^j(\vec{k})$ are the LCAO expansion coefficients.\\
In a chosen energy interval $\epsilon_{1} <\epsilon_j <\epsilon_{2}$ the local electronic density $\rho(\vec{r},\epsilon_1,\epsilon_2)$ is given by:
\begin{eqnarray}
\rho(\vec{r},\epsilon_{1},\epsilon_{2})=\Sigma_{\epsilon_j>\epsilon_{1}}^{\epsilon_j<\epsilon_{2}}\Sigma_{\vec{k}}\Psi^{*}_j(\vec{k},\vec{r})\Psi_j(\vec{k},\vec{r})
\end{eqnarray}
and one can subsequently compute the electronic contour plots to represent them in a chosen plane for a given energy interval.\\
Our sample contains 258 particles, confined in a
cubic cell of 19.21 \AA ~(the density corresponds thus to the experimental value of $2.75$~g/cm$^{3}$ \cite{boolchand}. At this density the pressure of the system is fairly small (close to 0.3~GPa) and does not affect the structural and vibrational properties of glassy GeS$_2$ as shown in previous studies \cite{blaineau2,blaineau1}). Periodic boundary conditions are applied in order to
limit surface effects, and only the $\Gamma$ point is used to sample the
Brillouin zone ($\vec{k}=\vec{0}$). At the beginning of the simulation a crystalline $\alpha$-GeS$_2$
configuration is melted during 60 ps (24000 timesteps), to obtain an equilibrated liquid system. The sample is then quenched through the glass transition (simulated
T$_g\approx 1200$ K) at a velocity of 6.8*10$^{14}$ K/s, and is finally relaxed at 300 K during 100 ps. Our results have been obtained from this equilibrated configuration, and the atomic charges have been averaged over the 40000 steps of the relaxation period.
\section{RESULTS}
	The Sankey-Niklewski scheme which has been described above allows the
	determination of the electronic energy eigenvalues of the glassy sample.
	The Electronic Density of States (EDOS) can be computed by binning these
	eigenvalues. We present the calculated EDOS in Fig.1, as well as the
	experimental valence spectrum which has been measured by X-ray
	Photoelectron Spectroscopy (XPS) \cite{foix}. Our simulation appears to be in good agreement with these experimental results, and presents three major contributions to the valence band, which will be called zone A [-16 eV,-12 eV], zone B [-9.8 eV,-6.5 eV] and zone C [-6.5 eV,-1.3 eV] respectively. The energy shift between experimental and theoretical EDOS \cite{foix} is relatively small in our work ($\approx$ 1 eV). The partial EDOS for $s$ and $p$~orbitals of germanium and sulfur atoms can be computed by summing the $|C_{\mu}^j(\vec{k})|^2$ for each element and each orbital.  We have scaled the partial EDOS to ensure that their sum is equal to the total EDOS, thereby including the interstitial contribution. The results are reported in Fig.2, where the solid line is the total EDOS, and the dashed area represents the contribution of a particular orbital.\\
We can clearly see that the zone A is mainly caused by the $3s$ orbitals of sulfur atoms, with a small contribution of the $4s$ orbitals of germanium atoms in the deepest part of the band [-16 eV,-14 eV]. The zone B shows a major contribution of the $4s$ orbitals of germanium, even though the presence of the $3s$ and $3p$ orbitals of sulfur is quite important (the respective concentration of the different elements must naturally be taken into account in this analysis). Finally, the band C appears to be mainly caused by $3p$ orbitals of sulfur atoms, with also an important contribution of the $4p$ orbitals of Ge. The highest energy zone of band C, which lies just below the Fermi level [-3.3 eV, -1.3 eV], is only caused by the $3p$ orbitals of the sulfur particles.\\
The analysis of the electronic structure of g-GeS$_2$ can be extended through the analysis of the contour plots, obtained by the calculation of $\rho(\vec{r},\epsilon_1,\epsilon_2)$ in the different energy zones [$\epsilon_1,\epsilon_2$], and localized around different structural arrangements.\\
We will first study the ''standard'' cases, in which the short-range coordination between particles is similar to what can be found in a crystalline structure, and we will subsequently focus on the localized electronic states caused by the bond defects found in our amorphous sample.
\subsection{Standard configurations}
The tetrahedral structure of g-GeS$_2$ makes it difficult to represent the contour plots in an intelligible 2D manner. The simplest way is to illustrate the electronic configuration of a 2-membered ring
(edge-sharing unit), since the four particles of these rings (2 Ge and 2 S) are almost located in the same plane, and in addition the electronic environment is in general the same than in the other configurations. In Fig.3 we show the contour plots of such entities in the upper and lower energy zones of bands A and B (the contour plots can slightly differ inside the same band). It can be seen in band A (Fig.3$(a),(b)$) that the signature of the 4s orbitals of the germanium atoms only appears in the lower energy region $(a)$ [-16 eV,-14 eV]. The link between the 3s orbitals of the two sulfur particles shows also a more bonding character in the lower region than in the upper region $(b)$ [-14 eV,-12 eV] of band A, where a nodal plane can be seen between the two S atoms (the S-S bond length in an edge-sharing unit is 3.3 \AA). The zone B (Fig.3$(c),(d)$) is predominantly caused by the $4s$ orbitals of Ge atoms, presenting $\sigma$ bonds with the $3s$ and $3p$ orbitals of the sulfur atoms. The lower region of band B [-9.8 eV,-7.6 eV] shows a more bonding character between Ge and S than the upper region [-7.6 eV, -6.5 eV], since the electronic density is more important between the two particles. On the contrary, the Ge-Ge interaction (d$_{Ge-Ge}$=2.9 \AA ~in an edge-sharing unit) is clearly anti-bonding in the lower region of zone B, where a nodal plane can clearly be seen.\\
In band C the results are different whether the intertetrahedral links are edge-sharing or corner-sharing. When a tetrahedron is not involved in a 2-membered ring, its intratetrahedral angle $\widehat{SGeS}$  fluctuates during the simulation between 100$^o$ and 115$^o$, because of the thermal vibrations of the system.
   Fig.4$(a)$ and 4$(b)$ show the
 contour plots of a S-Ge-S unit in band C at different times $t_1$ and $t_2$, when the
 $\widehat{SGeS}$ angle is respectively maximal and minimal. When the intratetrahedral angle is 
maximal (Fig.4$(a)$), the $p$ orbitals
of Ge and S atoms appear to be connected ''linearly'' by $\sigma$ links of $p$ orbitals,
each particle with its nearest-neighbors. This configuration becomes impossible
when the $\widehat{SGeS}$ angle decreases below a given value ($\approx
105^o$), because an anti-bonding overlap tends to appear between the two sulfur atoms. The
electronic configuration becomes then what can be seen in Fig.4$(b)$, where a
$\sigma$ link of the $3p$ orbitals of sulfur begins to appear between the two
second-neighboring S atoms (the S-S bond length is then 3.4 \AA). The apparent
contribution of the Ge atom in the contour plot is actually the signature of its
bonds with its two other nearest S neighbors, which are respectively above and
below the represented plane. The germanium atom is therefore not involved in the link with
the two sulfur atoms at these energies and for this configuration. At ambient
temperature the local electronic density in zone C fluctuates with time between these two environments, independently for each tetrahedron.\\
In edge-sharing units, the thermal vibrations are less pronounced, and the
intratetrahedral angle $\widehat{SGeS}$ is permanently inferior to 100$^o$. A
$\sigma$ bond between $p$ orbitals of germanium and sulfur atoms is therefore
not possible. Fig.5$(a)$, 5$(b)$ and 5$(c)$ show the different electronic configurations which have been found for an edge-sharing unit in the lower, middle and upper regions of band C (respectively [-6.5 eV, -4.6 eV], [-4.6 eV, -3.3 eV], [-3.3 eV, -1.3 eV]).  We can see in Fig.5$(a)$ the electronic configuration in the lower region, where each particle presents a $\sigma$ bond of $p$ orbitals with one of its nearest-neighbors, and a $\pi$ bond of $p$ orbitals with the other neighbor, creating two ''chains'' of $\sigma$ bonds, linked together by $\pi$ overlaps. The middle zone (Fig.$(b)$) presents the same kind of configuration than what has been seen in Fig.4$(b)$, with another additional germanium. The sulfur atoms present thus a weak $\sigma$ bond of $3p$ orbitals with each other, while the germanium atoms are involved in their connection with their other respective nearest S neighbors. Finally, the upper zone (Fig.5$(c)$), which corresponds to the highest peak in the EDOS just below the Fermi level, is distinctively caused by non-bonding lone-pair $3p$ orbitals of sulfur atoms, which is similar to the results obtained in $As_2S_3$ systems \cite{elliott}. No signature of the germanium atoms is found at these energies.\\
All these results correspond to standard configurations which can also be found in a crystalline structure. They are extremely similar to the results obtained by Foix $et~al.$ in cluster simulations of glassy systems, using an {\em ab initio} model within a Hartree-Fock method \cite{foix}. We have studied in a previous work the different bond defects caused by the broken chemical order of amorphous g-GeS$_2$ \cite{blaineau2}. These bond defects appear to have an important impact also on the EDOS.
\subsection{Bond defects}
The bond defects present in the amorphous sample may be responsible for very localized states in the EDOS.
As a measure of the localization of the electronic eigenstates, we define the Mulliken charge $q_i(E)$ at atomic site $i$ for the eigenstate of eigenvalue E \cite{mulliken}. We calculate then the Inverse Participation Ratio (IPR) as: \begin{eqnarray}
IPR(E)=\Sigma_{i=1}^N (q_i(E))^2 .
\end{eqnarray}
where $N$ is the total number of atoms of the system. A value of 1 means that the state is perfectly localized, and a uniformly extended state leads to a value of $1/N$. The total IPR is represented in Fig.6$(a)$, as well as the partial IPRs, obtained by summing over the Ge atoms (Fig.6$(b)$) and the S atoms (Fig.6$(c)$). We can determine the presence of five groups of ''localized'' states (IPR(E)$>$0.1).\\
 Two of them appear in zone A. The most intense, at -13.2 eV, is localized on non-bridging sulfur atoms (which are connected to only one germanium). These bond defects
have been studied previously \cite{blaineau2} and have been found to exist for 14.53$\%$ of the S atoms of
the system. This undercoordination of sulfur particles shows however no specific signature in the contour plots, which were found to be extremely similar to those illustrated in Fig.3.$(b)$. The other group of localized states, at -12 eV, is clearly localized on sulfur particles involved in homopolar bonds, which were found to appear for 3.48$\%$ of the S particles of the sample. The corresponding contour plot is shown in Fig.7$(a)$. We can see an anti-bonding $\sigma$* link between the two S particles (d$_{S-S}$=2.26 \AA), which is coherent with the analyzis done above in edge-sharing units between second-neighbor sulfur atoms (Fig.3$(b)$). It can be noted that these last states at -12~eV are slighlty separated from the rest of the states in zone A (Fig.6$(c)$).\\
The most localized states of band B, present at -9.6 eV
(Fig.6$(b)$), appear to be caused by the germanium atoms which form homopolar Ge-Ge bonds. These eigenstates are attributed to
$\sigma$ bonds between the $4s$ orbitals of the germanium atoms. Fig.7$(b)$ shows the corresponding contour plots around a Ge-Ge ethane-like unit, previously found to appear for 2.32$\%$ of the germanium atoms of the sample (d$_{Ge-Ge}$=2.42\AA) \cite{blaineau2}. It should be noted that the link between the germanium atoms is clearly bonding in this case, whereas the link between second Ge neighbors in edge-sharing units was strongly anti-bonding (cf. Fig.3.$(c)$).\\

The zone C presents two groups of localized states (IPR(E)$\approx$0.2) at -1.9 eV and -1.4 eV. The first one is clearly localized on sulfur atoms which are engaged in homopolar bonds. The free $3p$ orbitals of S atoms, previously attributed to this zone, interact with each other because of the small distance between the S particles (d$_{S-S}$=2.26~\AA). These localized states were also found to exist in $As_2S_3$ systems \cite{elliott}. The corresponding contour plot is shown in Fig.7$(c)$. We recognize an anti-bonding $\pi$* link between the $3p$ orbitals of the sulfur atoms, and two nodal planes can be seen. The S-S homopolar bonds can appear whether the sulfur atoms are bridging or non-bridging (connected to 2 or 1 Ge atoms). The localized eigenstates at -1.9 eV are caused by the non-bridging sulfur atoms, whereas the bridging S atoms are responsible for the second group of states, at -1.4 eV (Fig.6.$(c)$), and present the same contour plots as Fig.7$(c)$. These states are the last states below the Fermi level, which in a molecular description would correspond to the HOMO (Highest Occupied Molecular Orbital). It should be noted that the experimental EDOS obtained by XPS measurements presents, as well as our calculated EDOS, a small shoulder at the end of the valence band (Fig.1), which has not been found in the crystalline structures \cite{foix}. This shoulder could be an experimental signature of homopolar S-S bonds in g-GeS$_2$.
 We can also identify localized Ge states at these energies (Fig.6$(b)$), slightly below the HOMO. This localization on the germanium atoms is completely independent of the S-S homopolar bonds, and involves another group of particles presenting bond defects. These bond defects appear to be undercoordinated germanium atoms, which are connected to only 3 sulfur atoms instead of 4. The corresponding contour plots are shown in Fig.7$(d)$, in the plane containing the central Ge particle and two of its three nearest S neighbors. We recognize an anti-bonding $\sigma$* link between the $4s$ orbital of the germanium atom and the $3p$ orbitals of the sulfur particles. The third S neighbor, which is not in the plane, presents the same kind of overlap. It should be noted that no signature of the $4s$ orbitals of Ge has been found in this energy zone except in this kind of bond defects.
\\ In the study of the electronic properties of materials, the localized energy states at the tail of the conduction band are also of interest. The lowest localized (IPR(E)=0.33) energy state of the conduction band (called LUMO in a molecular description) was found at +1.5 eV (E$_F$=0), and we can see in Fig.8 that it is arising from anti-bonding $\sigma$* links between $3p$ orbitals of the same sulfur atoms that are responsible of the HOMO (i.e. those involved in S-S homopolar bonds). The optical gap is close to 3 eV, which is slightly inferior to the experimental value of 3.2 eV obtained by resonant Raman spectroscopy \cite{raman}. A second group of states, localized (IPR(E)=0.28) on 3-fold coordinated sulfur atoms, was found at an energy slightly superior to the energy of the LUMO (+1.8 eV). All these localized states near the band edges are crucial for the optical and photostructural properties of glassy GeS$_2$ and the knowledge of the above given characteristics is therefore essential.
\subsection{Atomic Charges}
It is well known that the charge of an atom cannot be measured by experiments,
and is therefore impossible to define uniquely. Nevertheless, there are
different methods such as L\"owdin \cite{lowdin} or Mulliken \cite{mulliken2} population analysis,
which can be interesting tools when comparing different configurations within the
{\em same} description. Our aim is to analyze the dependence of the local atomic charges
with the structural environment of the atoms, and in particular with the bond
defects, which have been previously studied \cite{blaineau2}. In this work we have chosen the L\"owdin
description. It should however be mentioned that the non-self-consistent Harris functional is known to slightly exaggerate the charge transfers between the atoms.
 The atomic charge q is calculated by the difference
between the amount of electrons of the neutral atom and the ''real'' amount of electrons in the given chemical environment (results are reported in Table I). The error bars correspond to the standard deviation of the atomic charges.\\
\begin{center}
\begin{tabular}{c|cc}
\multicolumn{3}{c}{TABLE I~~~~~Partial atomic charges as a function of the local structural environment}\\
\hline
{\bf Ge} environment& ~~~~~~~q   & Percentage  \\
\hline
{\bf Ge}(S$_4$)&~~~~~~~+0.94 $_-^+$0.02&95.36$\%$  \\

{\bf Ge}(S$_3$)&~~~~~~~+0.57 $_-^+$0.01& 1.16$\%$\\
{\bf Ge}(S$_2$)&~~~~~~~+061 $_-^+$0.01& 1.16$\%$\\
{\bf Ge}(GeS$_2$)&~~~~~~~+0.81 $_-^+$0.01& 2.32$\%$\\
\hline
\hline
{\bf S} environment&~~~~~~~q&Percentage\\
\hline
Ge-{\bf S}-Ge&~~~~~~~-0.46 $_-^+$ 0.05& 68$\%$\\

Ge-{\bf S}&~~~~~~~-1.07 $_-^+$0.03 &14.53$\%$ \\
{\large{$^{Ge}_{Ge}>$}}{\bf{S}}-Ge&~~~~~~~0 $_-^+$0.06& 13.95$\%$ \\
Ge-{\bf{S}}-S&~~~~~~~-0.17 $_-^+$ 0.03&2.32$\%$\\
{\large{$_{Ge}^{Ge}>$}}{\bf{S}}-S&~~~~~~~+0.36 $_-^+$0.04  &1.16$\%$\\
\hline
\end{tabular}
\end{center}
\vspace*{1cm}
As expected, the general polarity of the Ge-S bond is Ge$^+$-S$^-$. In an ''ordered'' 
configuration, in which each Ge has four S neighbors and each S has two Ge neighbors, the polarity appears to be +0.94 for Ge and -0.46 for S (it should be noted that the charge transfers obtained with a Mulliken population analysis were found to be slightly superior than those provided by the present L\"owdin description ($\approx$0.15), independently of the structural configuration. Foix $et~al.$ have found charges of +0.77 for Ge and -0.43 for S in cluster calculations of glassy systems, with a Hartree Fock {\em ab initio} model and a Mulliken description \cite{foix}. These results are therefore closer to our L\"owdin charges than those obtained by our Mulliken description). The charge on the germanium atoms decreases when its number of nearest sulfur neighbors decreases ($\approx$ +0.6). Nevertheless, in an ethane-like unit made of a homopolar Ge-Ge bond and in which each germanium has three S neighbors, the charge on both germanium atoms is only slightly lower (+0.81) than in a usual tetrahedral configuration.
\\
The charge on the sulfur atoms is more variable, and the corresponding error bars are thus more important than those attributed to the Ge atoms. When a sulfur is non-bridging (i.e. connected to only one Ge atom), its charge becomes extremely negative (-1.07). On the other hand, when a S atom is connected to three germanium, its charge becomes almost neutral. These configurations were found to exist respectively for 14.53$\%$ and 13.95$\%$ of the sulfur atoms in the sample \cite{blaineau2}.\\ 
 The S-S homopolar bonds present quite interesting results: It has been found \cite{blaineau2} that at least one of the two sulfur atoms involved in a homopolar bond must be non-bridging. In this case, the charge on the sulfur atoms appears to be quite small (-0.17). Nevertheless, when a S particle involved in a homopolar bond is bridging (i.e. it is connected to one S and two Ge atoms), its charge becomes positive (+0.36). It is therefore easy to understand why a configuration where two bridging sulfur atoms are bonded together is impossible, since it would produce an environment with 6 neighboring particles charged positively, which is energetically very unfavorable.\\
In order to make an in-depth analysis of these ($a~priori$ surprising) charge transfers we define here the short-range charge deviation $\Delta Q_{SR}(i)$ of a particle i, taking into account the atomic charge of that given particle as well as the charges on its nearest-neighbors (determined from the radial pair correlation function). The aim is to determine if the local charge fluctuations are balanced at short-range by the environment of the atoms, or if positive and negative zones can exist in the glassy sample. For each atom, we sum its L\"owdin charge with the weighted charges of its nearest-neighbors, where the weighting is performed over the number of nearest-neighbors of each particle. Thus, if an atom i is n(i)-fold coordinated, and its atomic charge is q(i), the short-range charge deviation will be calculated as follows: 
\begin{eqnarray}
\Delta Q_{SR} (i)=q(i)+\Sigma_{j=1}^{n(i)} \frac{q(j)}{n(j)}
\end{eqnarray}
In a crystalline structure, where no bond defects are present, this value is almost zero for all the particles of the system. We illustrate in Fig.9 the results obtained for our amorphous sample, in which the particles with a $\Delta Q_{SR}$ higher than +0.3 are represented in blue (black), and the particles with a $\Delta Q_{SR}$ lower than -0.3 are printed in green (grey). The remaining particles (which are in a neutral environment) appear in white in the figure. \\
It can clearly be seen that positive and negative zones containing more than one particle (a maximum of eight linked particles could be found) appear in the glassy sample. This is an additional electronic signature of the broken chemical order inherent to amorphous systems, and this local deviation from neutrality (up to nearest-neighbor distances) is directly caused by the bond defects present in the sample. Indeed only atoms with an environment different from the standard environment appear colored in Fig.9.

\section{CONCLUSIONS}
We have analyzed the electronic structure of an amorphous GeS$_2$ system through DFT based MD simulations. We have focused our study on the impact of the structural disorder on the electronic properties of these glasses. Our EDOS appears to be in perfect agreement with experiments, and the contour plots corresponding to standard configurations are similar to those obtained by cluster simulations presented by Foix $et~al.$ in crystalline structures with the use of a Hartree-Fock {\em ab initio} method. We have found three bands in the EDOS:\\
- The band A [-16 eV,-12 eV] is predominantly caused by the 3s orbitals of sulfur atoms, and presents localized states on non-bridging sulfur and on sulfur particles engaged in homopolar bonds.\\
- The band B [-9.8 eV,-6.5 eV] is attributed mainly to the 4s orbitals of germanium atoms, even though the contribution of 3s and 3p orbitals of sulfur atoms is non-negligible. We have found localized states involving Ge-Ge homopolar bonds in this energy region.\\
- The band C [-6.5 eV,-1.3 eV] contains the major signature of the Ge-S covalent bond, showing $\sigma$ and $\pi$ bonds between $p$ orbitals.
The highest energy zone of the valence band is due to free $p$ orbitals of sulfur atoms. Localized eigenstates involving 3-times coordinated Ge particles were found. The HOMO and the LUMO appear to be respectively caused by anti-$\pi ^*$ and anti-$\sigma ^*$ bonds of 3p orbitals of sulfur pairs linked by a S-S homopolar bond. The optical gap is close to 3 eV in our simulation (the experimental counterpart is 3.2 eV). \\
Finally, we have studied the local atomic charges, and their dependence on the structural environment of the atoms. The ''usual'' charge transfer in an ordered Ge(S$_4$)$_{1/2}$ configuration is +0.94 for the Ge atoms and -0.46 for the S atoms in our simulation, using a L\"owdin description within the Harris functional. These charges can be extremely variable for the sulfur atoms depending on their local structural arrangement, and can even become positive for the S atoms engaged in S-S homopolar bonds. We have determined that these local charge fluctuations are not balanced at short-range up to the nearest-neighbor environment, and are responsible of positively and negatively charged regions within the glassy sample.\\   

{\bf Acknowledgments}\\
We thank Jun Li for providing the computer code necessary to the computation of the electronic contour plots, and David Drabold for helpful discussions. Part of the numerical simulations were done at the ``Centre Informatique National de l'Enseignement Sup\'erieur'' (CINES) in Montpellier.\\

\hrule

\vspace*{0.3cm}
$^\star$ jund@lpmc.univ-montp2.fr
\vspace*{0.5cm}
\hrule
\vspace*{0.5cm}
\hspace*{-0.7cm}NB: We apologize for the bad resolution of some figures. This is not the case in the PRB version of the manuscript.

\centerline{\includegraphics[width=13cm]{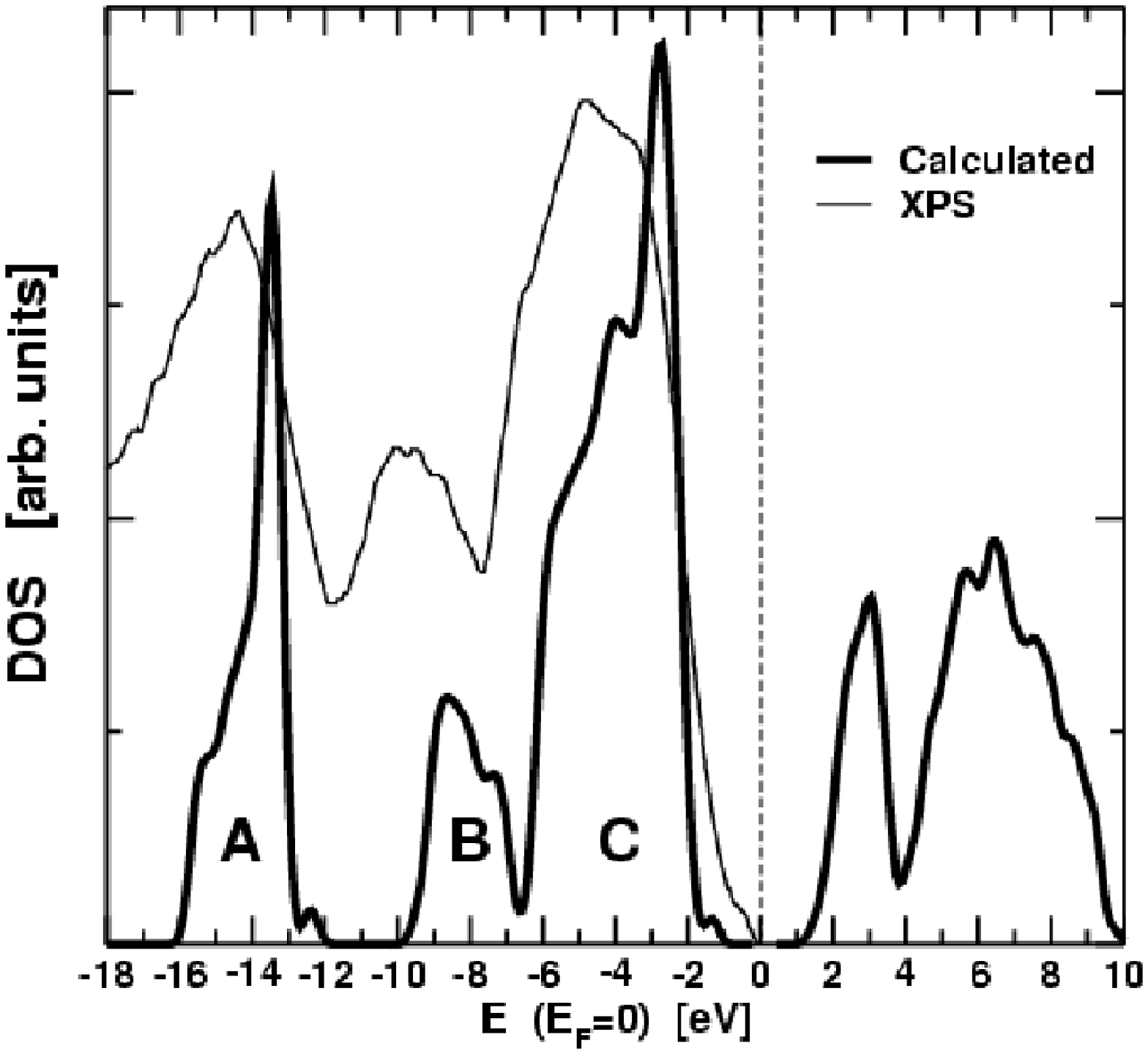}}
{\bf Figure 1.} Calculated EDOS and experimental valence spectrum obtained by 
XPS measurements {\cite{foix}}

\newpage
\centerline{\includegraphics[width=9cm]{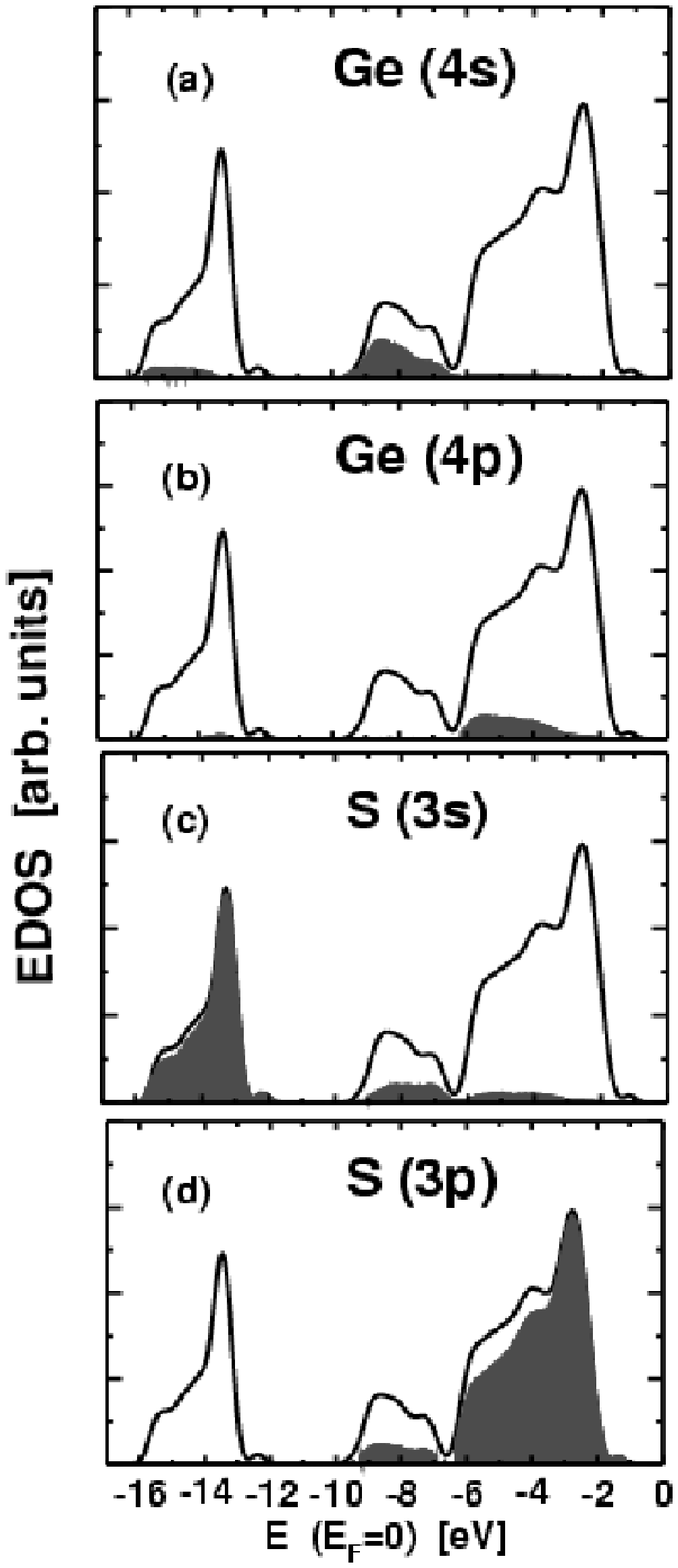}}
{\bf Figure 2.} Partial EDOS (shaded area) for the orbitals 4s$(a)$, 4p$(b)$ of germanium atoms, and 3s$(c)$, 3p$(d)$ of sulfur atoms and the total EDOS (solid line).
\newpage

\centerline{\includegraphics[width=7.5cm]{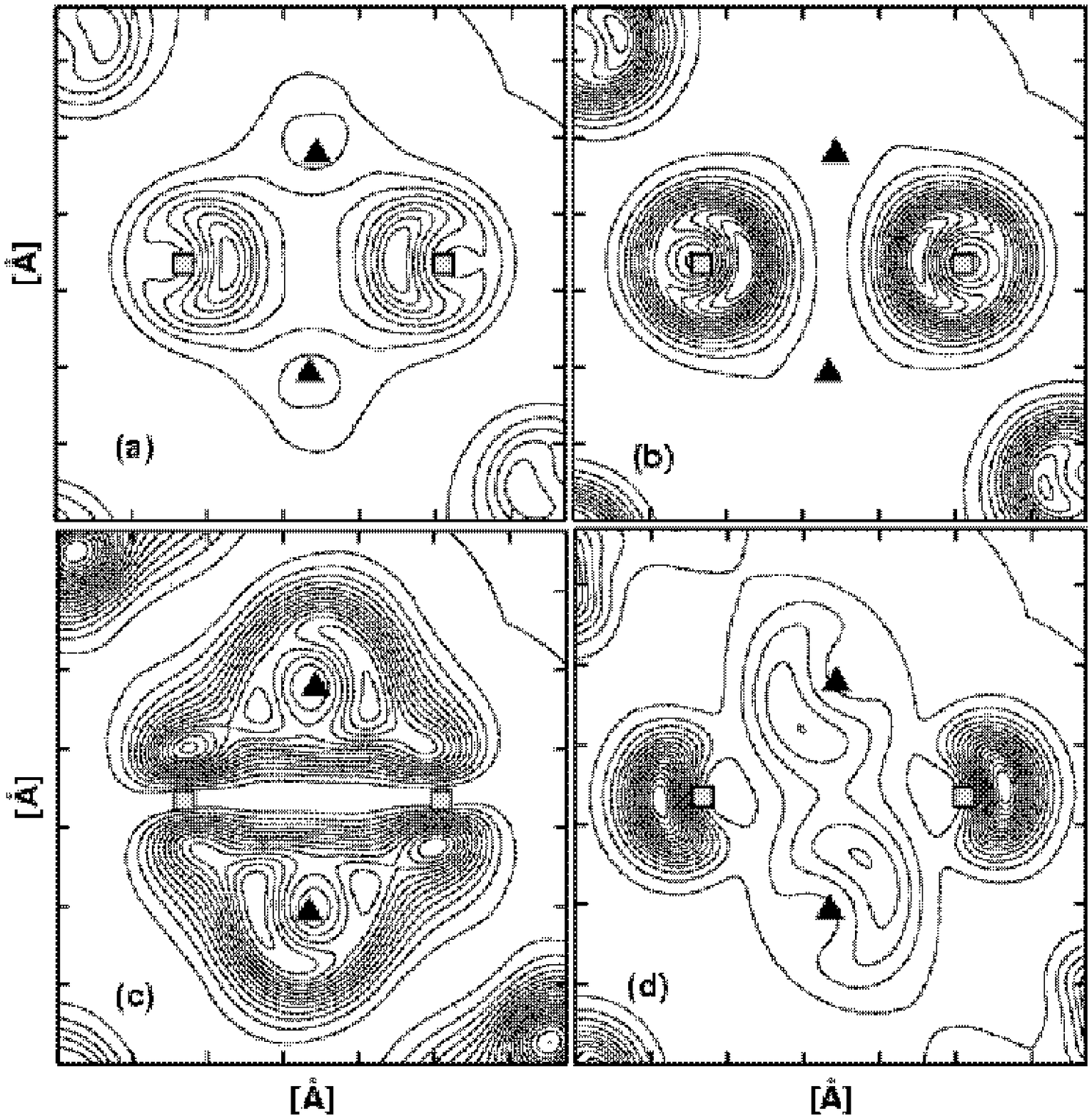}}
{\bf Figure 3.} Contour plots in an edge-sharing unit for band A in the lower $(a)$ and the upper $(b)$ regions (respectively [-16 eV,-14 eV] and [-14 eV,-12 eV], and for band B in the lower $(c)$ and the upper regions $(d)$ (respectively [-9.8 eV, -7.6 eV] and [-7.6 eV,-6.5 eV]). ($\blacktriangle$=Ge, $\square$=S) ($\rho(n+1)-\rho(n)$= 0.02(a,b), 0.008(c,d) e/\AA$^3$)

\centerline{\includegraphics[width=7.5cm]{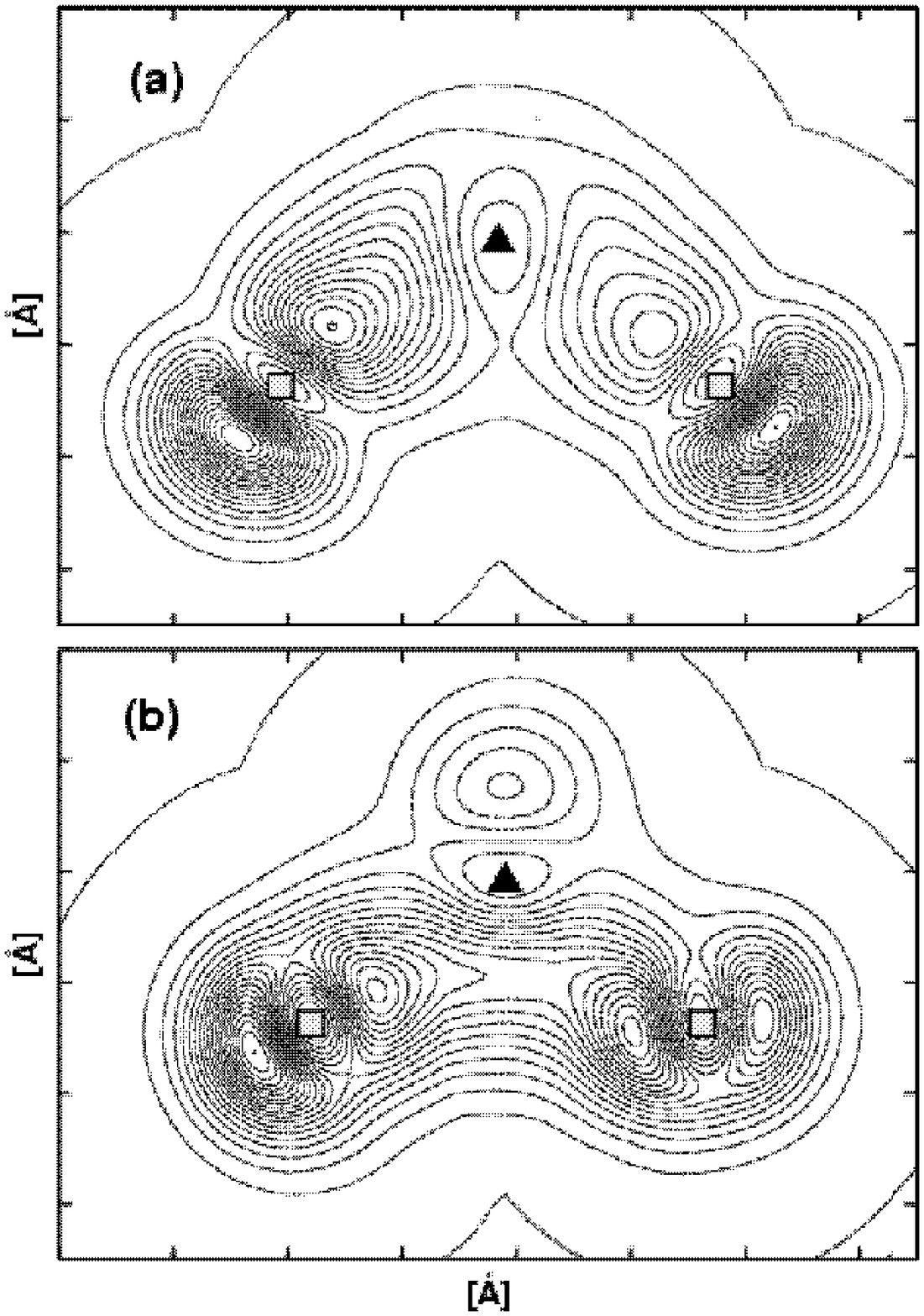}}
{\bf Figure 4.} Contour plots for band C in a tetrahedra not involved in an edge-sharing unit at times t$_1$ $(a)$ and t$_2$ $(b)$ when the intratetrahedral angle $\widehat{SGeS}$ is respectively maximal and minimal. ($\blacktriangle$=Ge, $\square$=S) ($\rho(n+1)-\rho(n)$= 0.098 (a,b) e/\AA$^3$)   
\newpage
\centerline{\includegraphics[width=11cm]{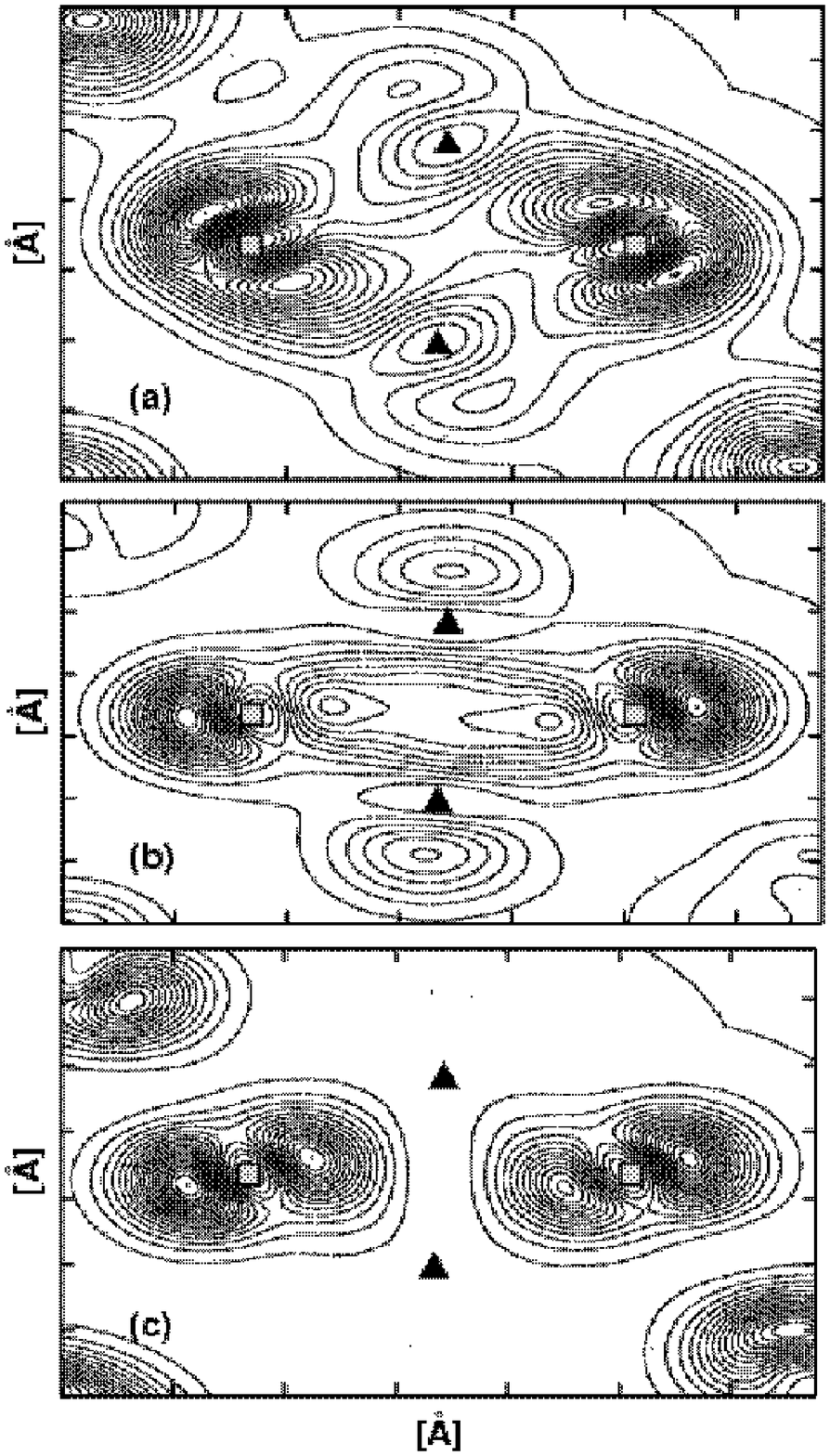}}
{\bf Figure 5.} Contour plots in an edge-sharing unit in the lower $(a)$, middle $(b)$ and upper $(c)$ zone of band C (respectively [-6.5 eV,-4.6 eV], [-4.6 eV,-3.3 eV] and [-3.3 eV, -1.3 eV]). ($\blacktriangle$=Ge, $\square$=S) ($\rho(n+1)-\rho(n)$= 0.013(a), 0.025(b), 0.034(c) e/\AA$^3$)

\centerline{\includegraphics[width=11cm]{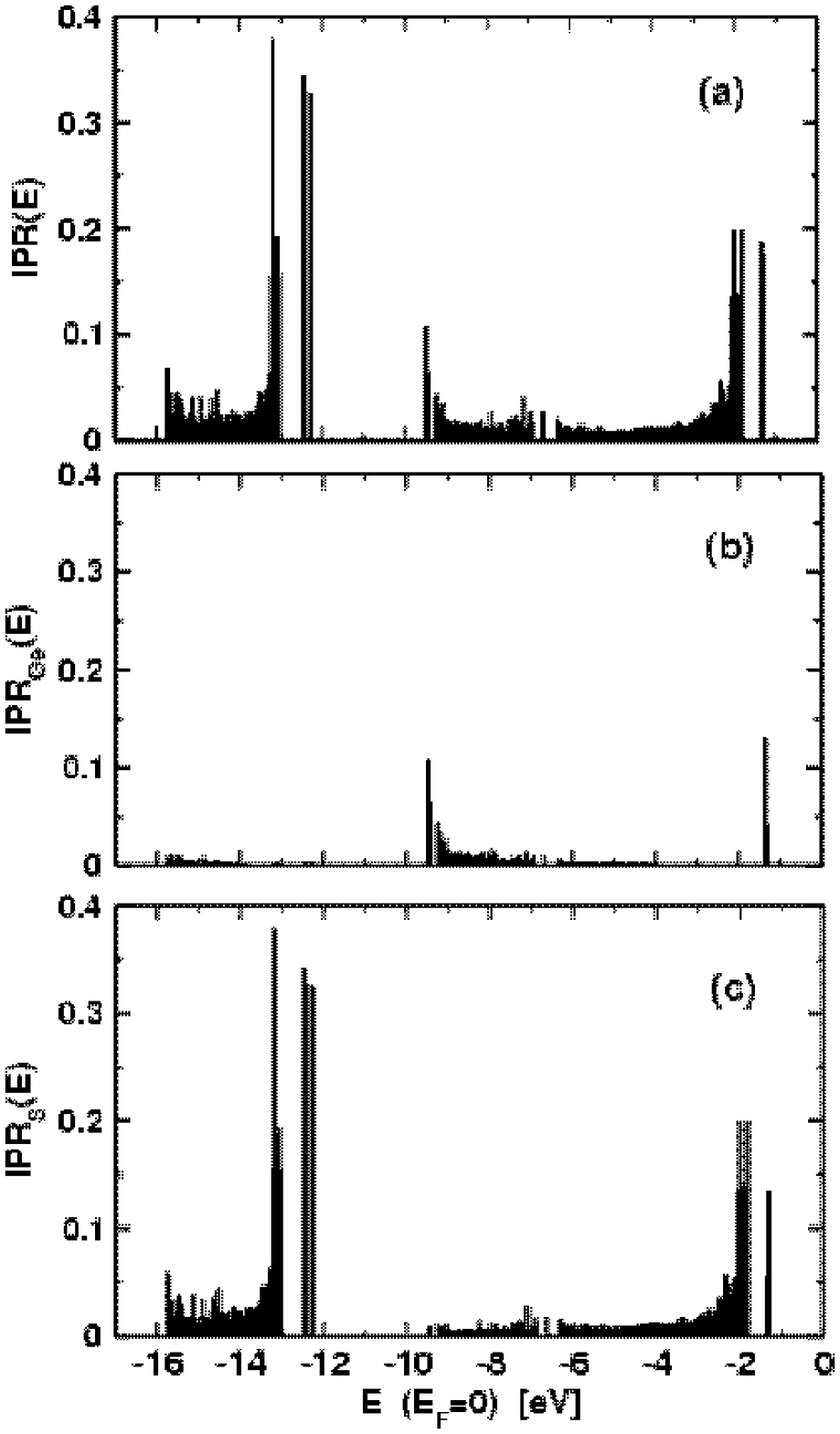}}
{\bf Figure 6.} Inverse Participation Ratio (IPR)  for the total system $(a)$ and for Ge$(b)$ and S$(c)$
\newpage
\centerline{\includegraphics[width=9.5cm]{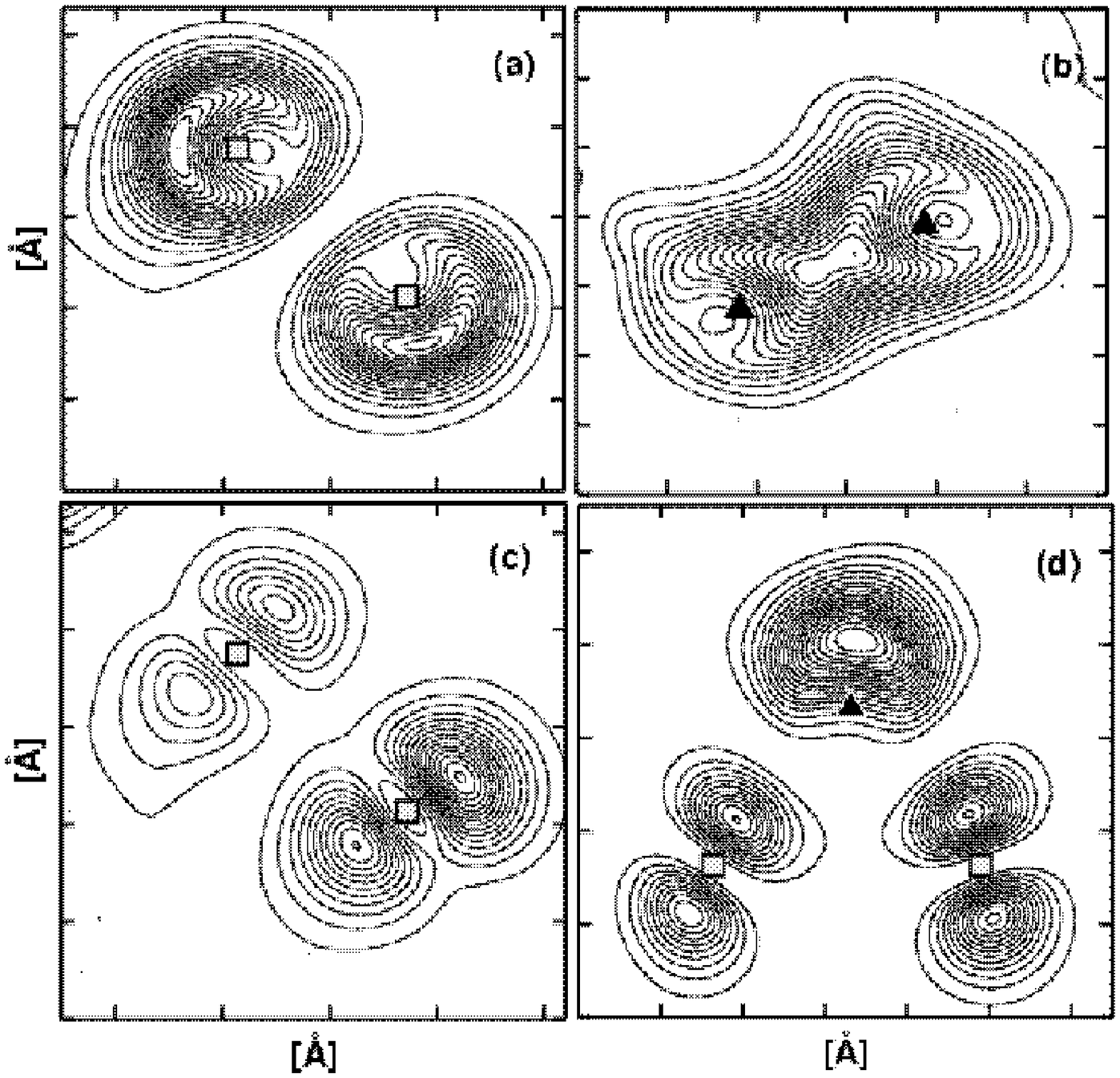}}
{\bf Figure 7.} Contour plots illustrating localized states at: $(a)$ -12 $^{+}_{-}0.1$ eV around a S-S homopolar bond; $(b)$ -9.6 $^{+}_{-}0.1$ eV around a Ge-Ge homopolar bond; $(c)$ [-1.9 eV,-1.4 eV] around a S-S homopolar bond; $(d)$ -1.4 $^{+}_{-}0.1$ eV around an undercoordinated Ge atom.($\blacktriangle$=Ge, $\square$=S) ($\rho(n+1)-\rho(n)$= 0.014(a), 0.003(b), 0.001(c), 0.002(d) e/\AA$^3$)

\centerline{\includegraphics[width=9cm]{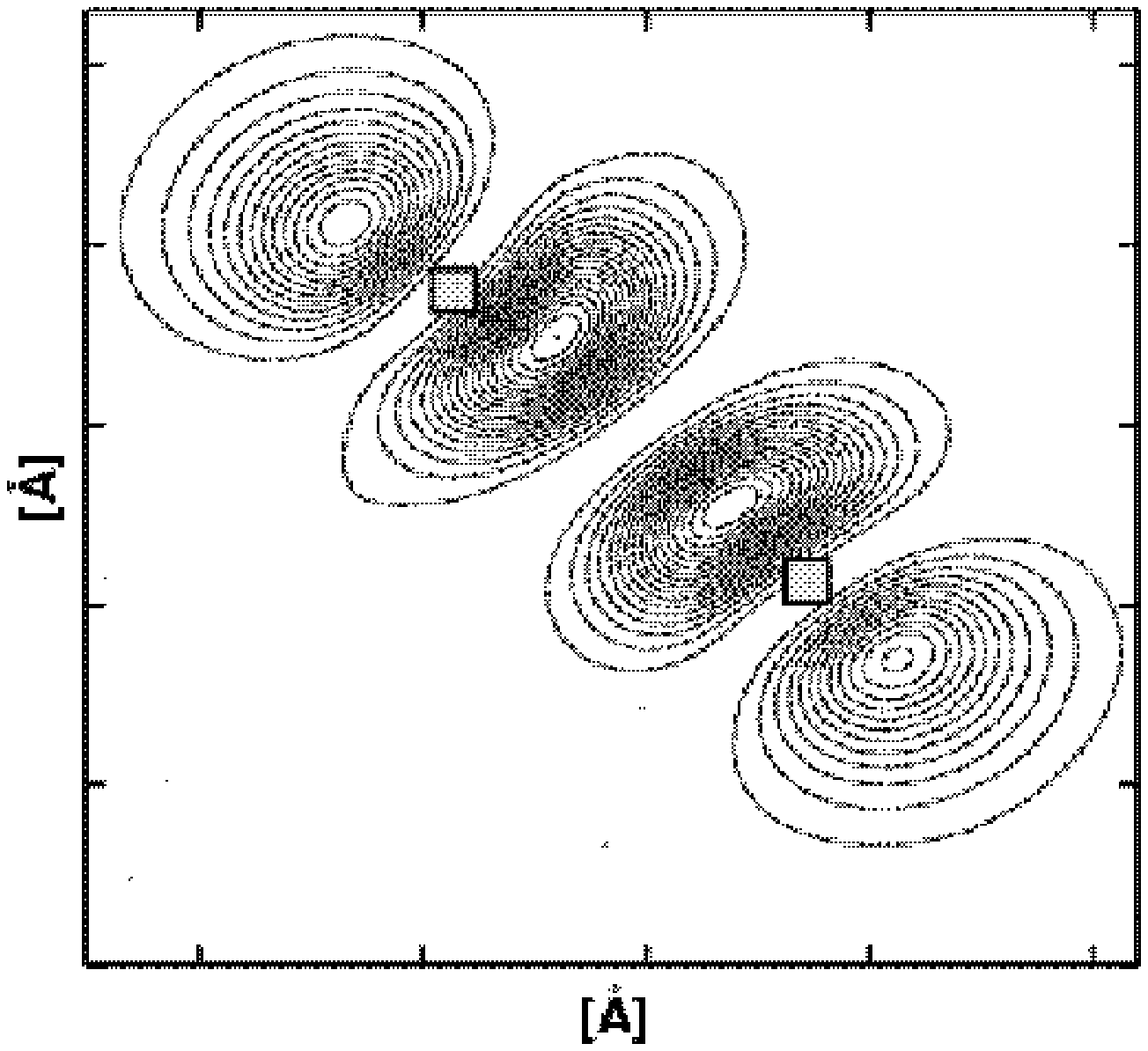}}
{\bf Figure 8.} Contour plot of the LUMO (+1.5 $^{+}_{-}0.1$ eV  ) localized around a S-S homopolar bond ($\rho(n+1)-\rho(n)$= 0.001 e/\AA$^3$)
\newpage
\vspace*{1cm}
\centerline{\includegraphics[width=14cm]{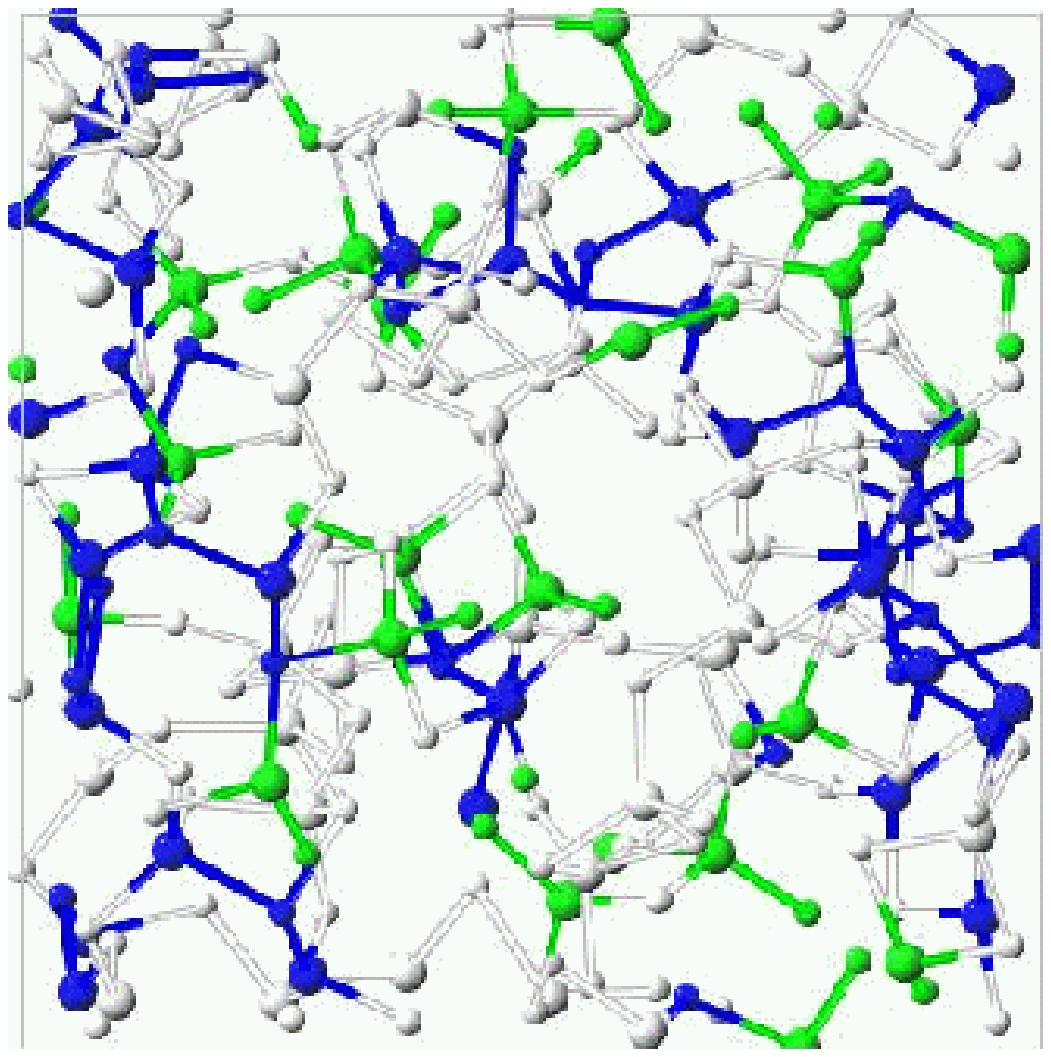}}
{\bf Figure 9.} (Color online) Short range charge deviation in our amorphous GeS$_2$ sample. For blue (black) particles, $\Delta Q_{SR} > +0.3$, and for green (grey) particles $\Delta Q_{SR} < -0.3$ (see text for definition)
\end{document}